\documentclass{article}

\begin{document}

\title{{\itshape Welcher Weg}?  A trajectory representation of a quantum Young's diffraction experiment}

\author{Edward R.\ Floyd \\
10 Jamaica Village Road, Coronado, CA 92118-3208, USA \\
floyd@mailaps.org}

\date{30 July 2007}

\maketitle

\begin{abstract}
The double slit problem is idealized by simplifying each slit by a point source.  A
composite reduced action for the two correlated point sources is developed. Contours
of the reduced action, trajectories and loci of transit times are developed in the
region near the two point sources.  The trajectory through any point in Euclidian
3-space also passes simultaneously through both point sources.
\end{abstract}

\bigskip

\footnotesize


\bigskip


\bigskip

\noindent Short running title:  {\it Welcher Weg}

\normalsize

\section{INTRODUCTION}

One of the first examples of wave-particle duality in quantum mechanics is the
double slit experiment which exhibits the wave and the particle properties.  The
wave properties are exhibited by diffraction patterns analogous to Young's optical
experiment while a detector still registers individual particles whose spatial
distributions are consistent with Young's diffraction. Attempts at detecting through
which slit the particle has passed destroys the interference between the two slits.

The trajectory representation has been developed as a deterministic theory of
quantum mechanics.$^{(\ref{bib:prd34}-\ref{bib:rc})}$ Faraggi and Matone have shown
that the foundations of quantum mechanics can be developed from the quantum
equivalence principle, which is consistent with the trajectory theory, without any
of the philosophy of the Copenhagen interpretation.$^{(\ref{bib:fm2})}$ The {\it
welcher Weg} aspect of the quantum Young's diffraction experiment is now ripe for an
investigation using the deterministic trajectory representation of quantum
mechanics.

The particular form, used herein, of the quantum Young's diffraction experiment
examines interference between a pair of coherent secondary point source activated
coherently by a primary source. Substituting secondary point sources for the
traditional slits simplifies the mathematics, which can be done in closed form,
without any loss of {\it welcher Weg} physics. Each secondary point source emits a
spherical wave that is a component of the total wave function for a solitary
diffracted quantum particle.  By themselves, the spherical wave components do not
represent particular states in the deterministic trajectory representation. The
total wave function is shown to be a dispherical wave function for the
self-entangled quantum particle that is synthesized from the pair of individual
spherical waves coherently emitted by the pair of secondary point sources.  The
terminology ``self-entangled" emphasizes that dispherical wave function may
represent a solitary quantum particle whose components, the pair of spherical waves
emitted by the coherent pair of secondary point sources, are entangled. A
self-entangled wave function for a quantum particle implies a solitary
self-entangled particle and is distinguished from an entangled wave function for an
entangled pair of quantum particles. This investigation develops from the
synthesized dispherical wave function the reduced action (Hamilton's characteristic
function) that is a generator of the motion for a solitary self-entangled quantum
particle. Subsequently the investigation, by applying Jacobi's theorem to the
reduced action, develops the trajectory and motion for a solitary self-entangled
quantum particle to resolve {\it welcher Weg}. This procedure is replicated to
establish various other trajectories using different constants of the motion. Both
cylindrical and prolate spheroidal coordinate systems are used in this exposition
for computational flexibility, as a computational check, and to gain insight. The
cylindrical coordinates are more familiar and more closely related to the
traditional presentation of the double slit experiment in the Fraunhofer region. The
prolate spheroidal coordinate system is the natural coordinate system for
investigating two-center phenomenon. Consequently, the results herein are presented
mostly in prolate spheroidal coordinates for heuristic purposes and to exhibit
insight.$^{(\ref{bib:mf})}$

The thrust of this investigation is to determine through which secondary point
source the trajectory passes.  The investigation therefore concentrates mainly in
the region near the pair of secondary point sources.  This region is well within the
region of Fresnel diffraction.  However, Fresnel approximations are not needed
herein as the diffraction for a pair of secondary point sources can be determined
exactly in closed form. All trajectories for the solitary self-entangled quantum
particle are shown to be strongly nonlocal as each individual trajectory is shown to
pass simultaneously through both secondary point sources through a series of the
trajectory segments that alternate forward and retrograde motion with respect to
time. The nonlocality of the dispherical particle implies that that the trajectory
is for a distributed particle rather than a point particle.

Manifestation of interference effects including reinforcement and destruction is
beyond the scope of this {\it welcher Weg} investigation. A companion article shows
how the deterministic trajectory representation exhibits interference effects
between plane wave functions that is consistent with the quantum equivalence
principle of Faraggi and Matone$^{(\ref{bib:fm2})}$ and does not resort to Born's
probability density of the Copenhagen interpretation.$^{(\ref{bib:irt})}$

Philippidis, Dewdney and Hiley have developed Bohmian trajectories for the double
slit experiment.$^{(\ref{bib:pdh})}$ Guantes, Sanz, Margalef-Roig and
Miret-Art\'{e}s have revisited the double slit experiment to develop Bohmian
trajectories, classical trajectories and the standard (wave function) quantum
representation for a ``soft-walled" double slit barrier.$^{(\ref{bib:guantes})}$
However, Bohmian mechanics$^{(\ref{bib:bohm})}$ and the trajectory representation
have different equations of
motion.$^{(\ref{bib:vigsym3},\ref{bib:fm2},\ref{bib:rc},\ref{bib:prd26})}$.
Consequently the trajectories of the two representations are different and imply
different physics and philosophy as one can learn by comparing the findings of these
investigations.

In Sect.\ 2, the wave function, generator of the motion (reduced action), trajectory
equation and equation of motion are developed for the solitary self-entangled
quantum particle of the quantum Young's diffraction experiment.  In Sect.\ 3,
examples of the contours of reduce action, trajectories and loci of transit times
are exhibited and discussed for the solitary self-entangled quantum particle. In
Sect.\ 4, the trajectory for the solitary self-entangled quantum particle is shown
to transit simultaneously both coherent secondary point sources resolving {\itshape
welcher Weg}. In the Appendix, the experiment is modified to render a set of
self-entangled wave functions that are shown to synthesize spherical waves.

\section{FORMULATION}

The formulation will be developed in both a modified prolate spheroidal coordinate
system $(\eta,\xi,\phi)$ and in the more familiar cylindrical coordinate system
$(\rho,z,\phi)$ to facilitate insight, accessibility and computational flexibility.
Computations were conducted in both coordinate systems, but the results will be
presented for heuristic purposes mostly in the modified prolate spheroidal
coordinate system with scale factors (metrical coefficients) modified by Morse and
Feshbach.$^{(\ref{bib:mf})}$  The two foci of the set of nested spheroids are at
$(\rho,z,\phi) = (0,\pm a/2,-\pi \le \phi \le \pi)$ in cylindrical coordinates. The
distances from the foci to a point $(\rho,z,\phi)$ are given in cylindrical and
prolate spheroidal coordinates by$^{(\ref{bib:mf})}$

\[
r_1 = [\rho^2+(z+a/2)^2]^{1/2} = a(\xi + \eta)/2 \ \ \mbox{and} \ \ r_2 =
[\rho^2+(z-a/2)^2]^{1/2} = a(\xi - \eta)/2.
\]

\noindent The modified prolate spheroidal coordinate system $(\xi,\eta,\phi)$, where
$\xi$ is the ellipsoidal coordinate, $\eta$ is the hyperboloidal coordinate, and
$\phi$ is the azimuthal coordinate, is specified by

\[
 \xi=(r_1+r_2)/a, \ \ \eta=(r_1-r_2)/a \ \ \mbox{and}\ \ \phi=\arctan (y/x). \
\]

\noindent   For the modified scale factors, see Morse and
Feshbach.$^{(\ref{bib:mf})}$  The foci of the spheroids are at $(\xi,\eta,\phi) =
(1,\pm 1,-\pi \le \phi \le \pi)$ in prolate spheroidal coordinates.  The line $\xi =
1$ lie on the principal axis of the set of nested prolate spheroids.

Let us consider two secondary point sources displaced from each other by the
distance $a$ where each secondary source specifies one of the foci for an infinite
set of spheroids. The secondary point sources are coherently actuated with equal
strength by a sole primary source so that the secondary sources emit simultaneously
components of equal magnitude of a solitary spinless quantum particle of mass $m$
and energy $\hbar^2k^2/(2m)$. Each of the two secondary point sources, if acting
alone, would be the source for a stationary spherical wave function given by

\begin{equation}
\psi_1 = \frac{\exp(i \mbox{\boldmath $k$}_1 \cdot \mbox{\boldmath $r$}_1)}{r_1} =
\frac{\exp(ikr_1)}{r_1} \label{eq:psi1}
\end{equation}

\noindent and

\begin{equation}
\psi_2  =  \frac{\exp(i \mbox{\boldmath $k$}_2 \cdot \mbox{\boldmath $r$}_2)}{r_2}
\label{eq:psi2} = \frac{\exp(ikr_2)}{r_2}
\end{equation}

\noindent where $\mbox{\boldmath $k$}_i$ is colinear with $\mbox{\boldmath $r$}_i$
and $\mbox{\boldmath $k$}_1 \cdot \mbox{\boldmath $k$}_1 = \mbox{\boldmath $k$}_2
\cdot \mbox{\boldmath $k$}_2 =k^2$. The two point secondary source problem is
azimuthally invariant. A dispherical wave function, $\psi_d$ can be synthesized from
its two components, $\psi_1$ and $\psi_2$, by$^{(\ref{bib:irt})}$

\begin{equation}
\psi_d = \psi_1 + \psi_2 = [r_1^{-2} + r_2^{-2} + 2r_1^{-1}r_2^{-1}
\cos(\mbox{\boldmath $k$}_1 \cdot \mbox{\boldmath $r$}_1 - \mbox{\boldmath $k$}_2
\cdot \mbox{\boldmath $r$}_2)]^{1/2} \exp \left[ i \arctan\left( \frac{r_2
\sin(\mbox{\boldmath $k$}_1 \cdot \mbox{\boldmath $r$}_1) + r_1 \sin(\mbox{\boldmath
$k$}_2 \cdot \mbox{\boldmath $r$}_2)}{r_2 \cos(\mbox{\boldmath $k$}_1 \cdot
\mbox{\boldmath $r$}_1) + r_1 \cos(\mbox{\boldmath $k$}_2 \cdot \mbox{\boldmath
$r$}_2)}\right)\right].   \label{eq:psid}
\end{equation}

\noindent  As $\psi_d$ is not factorable into a product of $\psi_1$ and $\psi_2$,
the two components, $\psi_1$ and $\psi_2$, are entangled in $\psi_d$.  The quantum
particle that is emitted from the two secondary point sources has $\psi_d$ as its
wave function and is self-entangled.  In this investigation, which examines the
behavior of $\psi_d$ as a solitary particle, the components $\psi_1$ and $\psi_2$ do
not represent a pair of particles that entangle with each other. For completeness,
the cosine term in the amplitude for $\psi_d$ in Eq.\ (\ref{eq:psid}) manifests
interference.

A generator of the motion, the reduced action (Hamilton's characteristic function),
$W_d$, for the self-entangled wave function can be extracted from Eq.\
(\ref{eq:psid}) as$^{(\ref{bib:irt})}$

\begin{eqnarray}
W_d & = & \hbar \arctan\left( \frac{r_2 \sin(\mbox{\boldmath $k$}_1 \cdot
\mbox{\boldmath $r$}_1) + r_1 \sin(\mbox{\boldmath $k$}_2 \cdot \mbox{\boldmath
$r$}_2)}{r_2 \cos(\mbox{\boldmath $k$}_1 \cdot \mbox{\boldmath
$r$}_1) + r_1 \cos(\mbox{\boldmath $k$}_2 \cdot \mbox{\boldmath $r$}_2)}\right) \nonumber \\
    & = & \hbar \arctan\left( \frac{[\rho^2+(z-a/2)^2]^{1/2} \sin[k_{\rho}\rho+k_z(z+a/2)] +
    [\rho^2+(z+a/2)^2]^{1/2} \sin[k_{\rho}\rho+k_z(z-a/2)]}{[\rho^2+(z-a/2)^2]^{1/2} \cos[k_{\rho}\rho+k_z(z+a/2)] +
    [\rho^2+(z+a/2)^2]^{1/2} \cos[k_{\rho}\rho+k_z(z-a/2)]}\right) \nonumber \\
    & = & \hbar \arctan\left( \frac{(\xi-\eta)\sin[k(\xi+\eta)a/2] +
    (\xi+\eta)\sin[k(\xi-\eta)a/2]} {(\xi-\eta)\cos[k(\xi+\eta)a/2] +
    (\xi+\eta)\cos[k(\xi-\eta)a/2]}\right).
\label{eq:wd}
\end{eqnarray}

\noindent This generator of the motion, $W_d$, for the dispherical particle of
quantum Young's diffraction experiment is in Euclidean 3-space and not in Hilbert
space. Faraggi and Matone have shown that the reduced action may be derived by the
quantum equivalence principle independent of the Schr\"{o}dinger
equation.$^{(\ref{bib:fm2})}$  The reduced action for the self-entangled wave
function is independent of $\phi$, which manifests azimuthal symmetry.  By Eq.\
(\ref{eq:wd}), $W_d$, albeit independent of $\phi$, does not contain a cyclic
coordinate. Neither is it further separable.  In the vicinity near one of the
secondary point source by Eq.\ (\ref{eq:wd}), the degree that $W_d$ mimics reduced
action of that secondary point source being the sole point source increases with
nearness to that point source.

The trajectory equation for the self-entangled wave function can be developed from
the reduced action by Jacobi's theorem $\beta_i = \partial W_d/\partial \alpha_i$
where $\alpha_i$ is one of the independent constants of integration and $\beta_i$ is
its associated constant coordinate.  This procedure here differs with Bohmian
mechanics, which does not employ Jacobi's theorem.$^{(\ref{bib:bohm})}$ Following
Goldstein$^{(\ref{bib:goldstein})}$ one may choose other independent quantities,
$\gamma_i$'s where each $\gamma_i$ is a function of all the $\alpha_i$'s. The
$\gamma_i$'s are constant momenta albeit not necessarily the integration constants
that arise by integrating the Hamilton-Jacobi equation. Jacobi's theorem still holds
for the $\gamma_i$'s. Here, $\hbar k_z$ has been selected for the representation in
cylindrical coordinates to be a constant momenta. The trajectory equation for
$\psi_d$ is rendered in the $\rho,z$-plane by Jacobi's theorem as

\begin{equation}
\beta_z = \hbar^{-1} \frac{\partial W_d}{\partial k_z} = \frac{r_2^2 [(z+a/2) -
\frac{k_z}{k_{\rho}}\rho] + r_1^2 [(z-a/2) - \frac{k_z}{k_{\rho}}\rho] + 2 r_1r_2
\cos[k(r_1-r_2)][z - \frac{k_z}{k_{\rho}}\rho]}{[r_2 \cos(kr_1) + r_1 \cos(kr_2)]^2
+ [r_2 \sin(kr_1) + r_1 \sin(kr_2)]^2} \label{eq:te1}
\end{equation}

\noindent where for compactness and didactic purposes not all the distances, $r_1$
and $r_2$, have not been expanded and where $k_{\rho}$ is another $\gamma$ given by
$k_{\rho} = +(k^2 - k_z^2)^{1/2}$ where the sign of $k_{\rho}$ is positive for
outgoing radiation.  We assume that the trajectory originates at the secondary point
source at the upper focus. Hence, for $r_2=0,\ \rho=0$ and $z=a/2$, then
$\beta_z=0$.  This permits us to simplify Eq.\ (\ref{eq:te1}) by

\begin{equation}
r_2^2 \underbrace{[(z+a/2) - \frac{k_z}{k_{\rho}}\rho]}_{\mbox{\scriptsize lower
source alone}} + r_1^2 \underbrace{[(z-a/2) -
\frac{k_z}{k_{\rho}}\rho]}_{\mbox{\scriptsize upper source alone}} + 2 r_1r_2
\underbrace{\cos[k(r_1-r_2)](z - \frac{k_z}{k_{\rho}}\rho)}_{\mbox{\scriptsize
interference effects}} =0. \label{eq:te2}
\end{equation}

\noindent  Equation (\ref{eq:te2}) has been organized into the weighted
contributions (sub-trajectories) from individual secondary sources acting alone and
from interference effects. The weighting of an individual contribution from a
secondary source acting alone is proportional to the square of the distance from the
alternate secondary source while the weighting of the contribution from the
interference effects is proportional to twice the product of the two distances from
the individual secondary sources. We note that these three sub-trajectories have the
same constant of the motion, $\eta_a$.  The corresponding trajectory equation
equation in prolate spheroidal coordinates may be expressed as

\begin{eqnarray}
(\xi-\eta)^2\{\eta_a[(\xi^2-1)(1-\eta^2)]^{1/2}-(1-\eta_a^2)^{1/2}(\xi \eta + 1)\} & \  & \nonumber \\
+ (\xi+\eta)^2\{\eta_a[(\xi^2-1)(1-\eta^2)]^{1/2}-(1-\eta_a^2)^{1/2}(\xi \eta - 1) & \  & \nonumber \\
+ 2(\xi^2-\eta^2) \cos(ka\eta)\{\eta_a[(\xi^2-1)(1-\eta^2)]^{1/2}-(1-\eta_a^2)^{1/2}\xi\eta\} & = & 0
\label{eq:te3}
\end{eqnarray}

\noindent where $\eta_a$ is the constant of the motion and is the $\eta$-asymptote
of the trajectory as it propagates without bound.  The top, middle and bottom lines
on the left side of Eq.\ (\ref{eq:te3}) represent in prolate spherical coordinates
the contributions due to the lower secondary source alone, upper secondary source
alone and interference effects respectively. The trajectory equations are implicit
functions. Equation (\ref{eq:te2}) renders $(\rho,z)$ mutually implicit while Eq.\
(\ref{eq:te3}) renders $(\xi,\eta)$ mutually implicit.

There is a fundamental simplicity that has been achieved by establishing the
trajectory of the quantum dispherical particle.  Only one constant of the motion is
needed to establish the trajectory even though the equation of motion may exhibit
vestiges of motion of its components and the interference between its components as
explicitly shown by Eq.\ (\ref{eq:te2}). Otherwise, had one worked directly with
trajectories for the component wave functions $\psi_1$ and $\psi_2$, then the worker
would need two constants of the motion to describe two trajectories (for $\psi_1$
and $\psi_2$) for the motion of the dispherical particle.

Note also that the trajectories determined by Jacobi's theorem are not necessarily
orthogonal to the contours of reduced action.$^{(\ref{bib:irt},\ref{bib:prd26})}$

For completeness, had we been investigating the entangled motion of two particles
where each secondary source had simultaneously emitted an identical particle, then
we would have proceeded as before and synthesized the entangled wave function, and
established the reduced action for the entangled pair.  The trajectory for the
entangled pair would be specified by a single constant of the motion determined by
Jacobi's theorem.  Note that this procedure would render the trajectory of the
entangled pair as a whole and not the trajectory of one particle of the entangled
pair under a quantum pressure due to a Bohmian quantum potential for the pair of
identical particles.  One can generalize for entangled ensembles of $N$ particles.
By extending a procedure given by Bohm,$^{(\ref{bib:bohm})}$ one can synthesize an
entangled wave function, $\psi_{\cal E}$, from the ensemble of wave functions,
$\psi_j, j=1,2, \cdots, N$, where each wave function represents one of entangled $N$
particles, by

\[
\psi_{\cal E} = ({\cal X}^2 + {\cal Y}^2)^{1/2} \exp[i \hbar \arctan({\cal Y}/{\cal X})]
\]

\noindent with

\[
{\cal X}  =  \Re \left[\sum_{j=1}^N \psi_j \right] = \sum_{j=1}^N \Re[\psi_j] \ \
\mbox{and} \ \ {\cal Y} = \Im \left[\sum_{j=1}^N \psi_j \right] = \sum_{j=1}^N
\Im[\psi_j].
\]

While this investigation to resolve {\itshape welcher Weg} concentrates upon the
Fresnel region ($1 \le \xi <6$ herein), let us now briefly examine qualitatively the
trajectories in the Fraunhofer region. In the limit $\xi \to \infty$, then by Eqs.\
(\ref{eq:te2}) and (\ref{eq:te3})  $z/\rho \to k_z/k_{\rho}$ rendering the expected
behavior of the trajectory in the infinitely outer region.  Hence, $k_{\rho}$ and
$k_z$ may be identified with the asymptotic direction of the trajectory. In prolate
spheroidal coordinates, the trajectory and the hyperboloid of revolution specified
by $\eta_a$ have the common asymptote which is a cone whose generating line from the
origin has the angle $\theta$ with the $z$-axis or the ellipsoidal principal axis
where $\theta$ is given, as expected, by

\[
\theta = \arccos (\eta_a) = \arctan(k_{\rho}/k_z).
\]

The concept that $\psi_d$ is synthesized from $\psi_1$ and $\psi_2$ follows from the
superpositional principal for linear homogeneous differential equations. In the
Appendix, a modified experiment is described where $\psi_1$ and $\psi_2$ are
synthesized from two different dispherical wave functions.

 The equation of motion for the dispherical wave function is rendered by
Jacobi's theorem, $t - \tau = \partial W_d/\partial E$ where $\tau$ specifies the
epoch.  Jacobi's theorem gives

\begin{equation}
t-\tau = \left( 1-\frac{2}{\xi^2+\eta^2+(\xi^2-\eta^2)\cos(k\eta a)} \right) \frac{m
\xi \eta a}{\hbar k \eta_a} \label{eq:eom}
\end{equation}

\noindent for motion projected across the $\eta$-coordinates. The equation of motion
must be consistent with the side relation that all points $(\xi,\eta)$ obey the
trajectory equation, Eq.\ (\ref{eq:te3}), with the constant of the motion $\eta_a$.
For $\eta_a \to 0$, Eq.\ (\ref{eq:eom}) becomes singular and may be replaced using
Eq.\ (\ref{eq:te3}) by an alternative form

\begin{equation}
t-\tau = \frac{m \eta a}{\hbar k (1-\eta_a^2)^{1/2}} \label{eq:eomxi}
\end{equation}

\noindent for motion projected across $\xi$-coordinates.

\section{APPLICATION}

\subsection{Reduced Action}

Let us now consider an example $m=1,\ \hbar=1,\ a=1,$ and $k=15.2$.  The value of
$k$ was chosen so that $ka/\pi$ would not be a rational number for greater
generality. The contours for reduced action for the dichromatic particle are
determined by Eq.\ (\ref{eq:wd}) for $W=0.5h,1h,1.5h,\cdots,5h$ and exhibited in the
$\xi,\eta$-plane on Fig. 1. By symmetry, Fig. 1 need only only cover the range $0
\le \eta \le 1$ as there exists a mirror symmetry on the plane $\eta=0$ in addition
to the azimuthal symmetry in $\phi$. The contours of constant $W_d$ orthogonally
intersect the plane $\eta=0$ manifesting the mirror symmetry and orthogonally
intersect the axis $\xi=1$ or $\eta=1$ manifesting azimuthal symmetry. The implicit
relationships between $\eta$ and $\xi$ or $z$ and $\rho$ were established by solving
Eq.\ (\ref{eq:wd}) numerically by the secant method. The contours of reduced action
manifest self interference on Fig. 1 as wrinkles in the contours reminiscent of the
serpentine contours exhibited for interfering plane waves in the companion
paper.$^{(\ref{bib:irt})}$. These wrinkles diminish as the ratio $r_2:r_1 \to 0$.
The locations of these wrinkles are shown by Fig. 2 to occur near  $ka\eta
=(2n-1)\pi,\ n=1,2,\cdots$ in general or, for $k=15.2$ and $a=1$, in the vicinity of
the hyperboloids $\eta = 0.207, 0.620$ near where maximum destructive interference
occurs between $\psi_1$ and $\psi_2$. The distribution of the wrinkles in $W_d$ with
regard to $ka\eta $ can be substantiated be considering Fig.\ 3 which exhibits the
contours for $W_d$ for $m=1,\ \hbar=1,\ a=1,$ and $k=24.3$.  This change in $k$
induces four wrinkles in Fig.\ 3 in the vicinity of the hyperboloids $\eta = 0.129,
0.388, 0.646, 0.905$ where maximum destructive interference occurs.

The behavior of the contour of $W_d$ at the origin $(\xi,\eta)=(1,0)$ is interesting
and exhibited in Fig.\ 4. Figure 4 exhibits the behavior of three contours separated
by $0.001h$ in action in the vicinity of the origin. Figure 4 is also presented in
cylindrical coordinates as prolate spheroidal coordinates do not render any physical
insight at the fine scale used therein.  Contours of constant reduced action less
than approximately $W_d=1.316815$ are disjointed with two sets of contours: each
enclosing one or the two secondary point sources. At approximately $W_d=1.316815$,
the two disjointed contours of reduced action merge at the origin. At the origin,
the contour must instantaneously transition from orthogonal to the ellipsoidal
principal axis $(1,\epsilon)$ to orthogonal to the plane $(1+\epsilon,0)$ in the
limit $\epsilon \to 0$. Likewise a similar situation is happening for the
mirror-symmetric partner contour orthogonal to the ellipsoidal principal axis
$(1,-\epsilon)$.  As $\epsilon \to 0$ from above, the radii of curvature of the two
mirror-symmetric partner contours are vertically aligned and go to zero from
opposite signs, the mirror-symmetric partner contours become tangent albeit the
symmetric partner contours have infinite curvature of opposite signs inducing a zero
degree of osculation, and their evolutes merge vertically at $(1,0)$.   At and only
at $\epsilon = 0$, the infinite curvatures of opposite sign at the point of tangency
$(1,0)$ of the two mirror-symmetric contours form a third contour from the two
mirror-symmetric partner contours.  This third contour has infinite curvature at
$(1,0)$ but whose radius of curvature while zero is, in the limit $\epsilon \to 0$,
horizontally aligned, which allows the two mirror symmetric partner contours to be
joined at $\epsilon = 0$ to form a single contour. Succeeding contours of larger
action higher will be orthogonal to the plane $\eta=0$.

\subsection{Trajectories}

We consider the same $m=1,\ \hbar=1,\ a=1,$ and $k=15.2$ for trajectories that we
used for investigating reduced action. The initial trajectory for examination leaves
the upper secondary source with the prolate spheroidal constant of motion $\eta_a=
-\sin(\pi/18)$, which has been chosen to be explicitly negative so that the
secondary point source and $\eta_a$ are in opposite hemispheres with regard to the
sign of $\eta$ . The corresponding cylindrical constant of the motion is with $\hbar
k_z=-\hbar k\sin(\pi/18)$.  The trajectory equations, Eqs.\ (\ref{eq:te2}) and
(\ref{eq:te3}), are implicit and solved numerically by the secant method in
cylindrical and prolate spheroidal coordinates respectively. The trajectories in
prolate spheroidal coordinates are a monotonic function of $\eta$ while in
cylindrical coordinates neither $\rho$ nor $z$ are the trajectories monotonic.
Hence, the numerical process was better behaved in prolate spheroidal coordinates
with regard to convergence by the secant method for successive points of the
trajectory. The resulting trajectory transits between the two secondary point
sources as exhibited in Fig.\ 5.  For this reason, this trajectory is called a
``confined" trajectory.  As the double point source experiment has azimuthal
symmetry, Fig.\ 5 exhibits the trajectory projected onto the $\rho,z$-plane in
cylindrical coordinates.  Figure 5 is presented in cylindrical coordinates to
accommodate a change of scale by a factor of ten in $\rho$ at $\rho=0.1$ to
facilitate exposition of significant detail and the entire trajectory between the
two secondary point sources. Nevertheless, prolate spherical coordinates still
renders better insight when examining the trajectory presented on Fig.\ 5.

Figure 5 is misleading as the apparent symmetry of the trajectory about the $\eta
=0$ or $z =0$ plane in Euclidean 3-space is spurious. In prolate spheroidal
coordinates, $\xi \ge 1$ while in cylindrical coordinates $\rho \ge 0$. The point on
the trajectory at the origin $(\xi,\eta)=(1,0)$ or $(\rho,z)=(0,0)$, is an inflexion
point in Euclidean 3-space. The trajectory mimics a cubic equation in the
neighborhood of the origin where the symmetric pair of constant reduced action
contours osculate with zero curvature.  As the trajectory transits $z=0$, $\phi$
changes value by $\pm \pi$. This inflexion point  of the trajectory in Euclidean
3-space at the coordinate origin $(\rho,z)=(0,0)$ induces the trajectory to be
antisymmetric in Euclidean 3-space about the plane $\eta=0$ or $z=0$.  The
trajectories intersect the contours of constant reduced action in opposite direction
in the two hemispheres.

The trajectory, as exhibited by Fig.\ 5, has turning points in $\xi$ in the vicinity
of $ka\eta = \pm 2\pi,\pm 4\pi$ or in the vicinity of $\eta = \pm 0.413, \pm 0.827$
or $z = \pm 0.207, \pm 0.413$. At these points, the interference between $\psi_1$
and $\psi_2$ reinforce each other. This is analogous to the lower turning points for
the example considered in the companion paper.$^{(\ref{bib:irt})}$ Other turning
points in $\xi$ occur for some $\xi > 1$ in the vicinity of $ka\eta = \pm \pi, \pm
3\pi$ or in the vicinity of the hyperboloids $\eta = \pm 0.207, \pm 0.620$ where the
interference between $\psi_1$ and $\psi_2$ oppose each other  as the $\cos(ka\eta)$
term in Eq.\ (\ref{eq:te3}) is the super preponderate cause of destructive
interference. These turning points are analogous to the upper turning points for the
example considered in the companion paper.$^{(\ref{bib:irt})}$

There remains two turning points in $\eta$, which manifest local maximum destructive
interference between $\psi_1$ and $\psi_2$, on Fig.\ 5 that are located
approximately at points $(\xi,\eta) \approx (1.0065,\pm 0.954)$ or $(\rho,z) \approx
(0.0171,\pm 0.480)$.  Here, the contribution of the factor $\cos(ka\eta)$ in Eq.\
(\ref{eq:te3}) contributing to self-interference is no longer super preponderate
contribution near a secondary source point $(\xi,\eta)=(1,\pm1)$. The choice of $ka
\ne 2n\pi, n=1,2,3,\cdots$ preempts the existence of latent next ``regular" turning
point at $ka\eta = \pm 5\pi$ for such $|\eta| \not\le 1$ would be nonphysical.
Nevertheless, $\psi_d$, which has no self-interference at either secondary point
source, immediately acquires self-interference upon sortieing from either secondary
point source that increases to leading order as $(\xi-\eta)(\xi+\eta)$ in the
vicinity of the secondary point sources that in turn induces these ``irregular"
turning points in conjunction with the behavior of $\cos(ka\eta)$ factor in Eq.\
(\ref{eq:te3}).

Let us now investigate a set of selected ``confined" trajectories that leave the
upper secondary point source with various values of $\eta_a \le 0$  so that the
secondary source and $\eta_a$ are in opposite hemispheres . Figure 6 exhibits the
set of selected trajectories whose constants of the motion are given by

\[
\eta_a = -\sin(0), -\sin(\pi/8), -\sin(\pi/4), -\sin(3\pi/8) \approx
-0,-0.383,-0.707,-0.924,
\]

\noindent  where $-0$ denotes that the limit $\eta_a \to 0$ is from below.  By
symmetry, Fig.\ 6 need exhibit only the right upper quadrant. A fifth trajectory for
constant of the motion, $\eta_a=-\sin(\pi/2)=-1$ superimposes line $\xi=1$ on Fig.\
6 in the range $0 \le \eta \le 1$. The trajectories for $\eta_a \approx
-0.383,-0.707,-0.924$ cross at their mutual inflexion point in Euclidean 3-space at
the origin $(\xi,\eta)=(1,0)$.  For $\eta_a<0$, in the limit that the constant of
the motion $\eta_a \to 0$ from below, then its trajectory too goes through the
origin and crosses those other trajectories with $\eta_a < 0$ there.  Thus, the
origin, $(1,0)$, is a focus for ``confined" trajectories.

The trajectories are mutually tangent at the turning points at $\xi=1$ and $ka\eta =
2\pi, 4\pi$ where there is maximum reinforcement between $\psi_1$ and $\psi_2$. The
trajectories have common turning points at $(\xi,\eta) = (1,2\pi/ka),(1,4\pi/ka)
\approx (1,0.413),(1,0.827)$  that form foci of the ``confined" trajectories on the
ellipsoidal principal axis $\xi=1$  The trajectories also do not cross at these
foci.

The trajectories have ``regular" turning points, where the values of $\xi$ attain
local maxima, near the hyperboloids $\eta = \pi/ka, 3\pi/ka \approx 0.207, 0.620$.
These ``regular" turning points are located at local maxima in the destructive
interference between $\psi_1$ and $\psi_2$. Due to the scale of Fig.\ 6, only the
``regular" turning point manifesting maximum destructive interference for the
trajectory with constant of the motion $\eta_a = -\sin(3\pi/8) \approx -0.917$ is
exhibited on Fig.\ 6 near the unexhibited hyperboloid $\eta = 3\pi/ka \approx
0.620$. For the selected family of exhibited trajectories, the other ``regular"
turning points of maximum destructive interference are displaced well off the scale
of Fig.\ 6.

The ``irregular" turning points exist on Fig.\ 6 where the trajectories attain $\xi$
values of local maxima for $3\pi/ka  \eta <1$.  These ``irregular" turning points
have values of $\eta$ between approximately 0.955 and 0.975 and increase as the
constant of the motion, $\eta_a$ increases.

A trajectory in Fig.\ 6 passes through an alternating series of turning points.  The
turning points where maximum destructive interference in $\psi_d$ occur are
interspersed with turning points (foci) where maximum reinforcement occurs. This
alternating series of turning points is reminiscent of the companion
paper$^{(\ref{bib:irt})}$ where alternating turning points manifest creation and
annihilation of trajectories.  At the turning points of maximum destructive
interference one trajectory segment in forward motion merges with a retrograde
trajectory for mutual annihilation while foci create a forward and retrograde
trajectory segments.  Thus, the trajectories of Fig.\ 6 have pattern of alternating
forward and retrograde segments with respect to the ellipsoidal coordinate $\xi$ (in
Section 3.3 the retrograde motion is shown to be also with respect to time).

Let us now investigate a set of selected trajectories with various values of $\eta_a
\ge 0$ so that the secondary point source and $\eta_a$ are in the same hemisphere
with regard to the sign of $\eta$. We first examine the trajectory that leaves the
upper secondary source with the spheroidal constant of the motion $\eta_a =
\sin(\pi/32) \approx 0.0980$. The corresponding cylindrical constant of motion
$\hbar k_z = \hbar k \sin(\pi/32) \approx 15.12681 \hbar$. This trajectory is
exhibited on Fig.\ 7 and is monotonically decreasing in $\eta$ as it asymptotically
approaches its constant of the motion $\eta_a$ where $\xi$ increases without bound.
As such, this trajectory is called ``free". The trajectory exhibits turning points
for local extrema in the value of $\xi$. The two turning points at local minima of
$\xi$ manifesting maximum reinforcement are located near the unexhibited
hyperboloids $\eta= 2\pi/ka, 4\pi/ka$ with values of $\xi>1$. In contradistinction
to the ``confined" trajectories, ``free" trajectories do not have foci on the
ellipsoidal principal axis $\xi=1$. The trajectory has turning points at local
maxima of $\xi$ near the unexhibited hyperboloids $\eta = \pi/ka, 3\pi/ka$ that are
well displaced off Fig.\ 7. The trajectory has an additional turning point near
$(\xi,\eta)=(3.409,0.128)$ after which the trajectory increases in $\xi$ without
bound and sharing the mutual asymptote with the hyperboloid $\eta = \sin(\pi/32)$.

Figure 8 exhibits the set of four selected ``free" trajectories exhibited whose
constants of the motion are given by

\[
\eta_a = +\sin(0), +\sin(\pi/8), +\sin(\pi/4), +\sin(3\pi/8) \approx
+0,+0.383,+0.707,+0.924
\]

\noindent  where $+0$ denotes that the limit $\eta_a \to 0$ is from above. The
secondary source and $\eta_a$ are in the same hemisphere for these trajectories. The
trajectories exhibited on Fig.\ 8 have positive values of $\eta_a$ while those for
Fig.\ 6 have negative. By symmetry, Fig.\ 8 need exhibit only the right upper
quadrant. A fifth trajectory with the constant of the motion, $\eta_a = +1$
superimposes on the line $\eta =1$ on Fig.\ 8 in the range $1 \le \xi $, that is the
the $z$-axis above the upper secondary point source in cylindrical coordinates. All
trajectories have monotonically decreasing $\eta$'s except for the one specified by
$\eta_a = +1$ as previously discussed. Each trajectory asymptotically approaches its
constant of the motion, $\eta_a$. Trajectories with constant of the motion $\eta_a >
3\pi/k$, that, for the trajectories exhibited on Fig.\ 8, includes the two
trajectories with $\eta_a \approx +0.924,+0.707$ respectively, do not ever reach the
value of $\eta \approx +0.620$ where $ \cos(ka\eta) = -1$ in Eq.\ (\ref{eq:te3})
manifesting latent maximum interference. This explains why the behavior of the
``free" trajectories become smoother with increasing $|\eta_a|$, (also true for
``confined" trajectories where in the limit $\eta_a \to |1|$ the confined trajectory
goes to the ellipsoidal principal axis, $\xi = 1$). The trajectory with $\eta_a=+0$
at $\eta=0$ must now be examined in the limit that $\eta_a \to 0$ from above. The
trajectory with constant of the motion $\eta_a = \sin(\pi/32)$ as exhibited on Fig.\
7 has, as noted in the previous paragraph, a turning point near
$(\xi,\eta)=(3.409,0.128)$ after which it proceeds asymptotically to its $\eta_a$.
This tuning point moves to the origin $(\xi,\eta)=(1,0)$ as $\eta_a \to 0$ from
above, and the trajectory for $\eta_a=0$ then superimposes on the line $\eta=0$ for
$\xi \ge 1$ on Fig.\ 8. Note that the behavior of the trajectory for in the limit
$\eta_a \to 0$ depends on whether the limit is approached from above (i.e., $\eta_a$
positive) or below (i.e., $\eta_a$ negative) where it was shown earlier in
conjunction with Fig.\ 6 that in the limit $\eta_a \to 0$ from below induces the
trajectory not to have a turning point at the origin.  The different behaviors for
the trajectories for $\eta_a=\pm 0$ is the reason for making explicit whether the
limit $\eta_a \to 0$ is taken from above or below.  The trajectories with $\eta_a =
+0,-0$ superimpose on each other for the segment between between $(\xi,\eta)=(1,1)$
and $(\xi,\eta)=(1,0)$, but differ beyond the origin. The trajectory for $\eta_a=+0$
then propagates out as a straight line, $(\xi,\eta)=(\xi \ge 1,0)$.  On the other
hand, the trajectory for $\eta_a=-0$ proceeds antisymmetrically, as previously
noted, to the lower secondary point source, $(\xi,\eta)=(1,-1)$. For completeness,
there is a trajectory originating from the lower secondary point source that is the
symmetric equivalent to the trajectory originating from the upper secondary point
source with $\eta_a=+0$.

By Fig.\ 6 the set of all ``confined" trajectories originating from the upper
secondary point source with a constant of the motion $\eta_a < 0$ lie on one side by
the trajectory for $\eta_a=-0$. By Fig.\ 8 the set of all ``free" trajectories
originating from the upper turning point with a constant of the motion $\eta_a > 0$
lie on the other side by the trajectory for $\eta_a=+0$.  By symmetry, an analogous
situation occurs for trajectories originating from the lower secondary point source
except that the trajectories with $\eta_a > 0$ are ``confined" while those with
$\eta_a < 0$ are ``free". A ``free" trajectory has its secondary point source and
constant of the motion, $\eta_a$ in the same hemisphere with regard to the sign of
$\eta$; a ``confined" trajectory, opposite hemispheres. The particular case of
$\eta_a=0$ depends on how $\lim _{\eta_a \to 0}$ is taken. The set of all
``confined" trajectories for which $\eta_a \ne 0$ form an open domain in
$\xi,\eta$-plane that is bounded by the trajectory originating from the upper
secondary point source with for $\eta_a=-0$ The set of all ``free" trajectories
originating from either secondary source combined with the trajectory $\eta_a=+0$
originating from the upper secondary source and its symmetric equivalent from the
lower secondary source form a closed domain in the $\xi,\eta$-plane that is the
compliment in the $\xi,\eta$-plane of the open domain formed by the open set of all
``confined" domains. Revolving the $\xi,\eta$-plane azimuthally through $2\pi$ in
$\phi$ shows that the trajectories for a quantum dispherical particle span Euclidean
3-space.

\subsection{Loci of Transit Times}

We consider the same $m=1,\ \hbar=1,\ a=1,$ and $k=15.2$  that we used for
investigating reduced action and trajectories for investigating the loci of transit
times for transits in the near region to assist resolving {\itshape welcher Weg}. We
have computed from Eqs.\ (\ref{eq:eom}) and (\ref{eq:eomxi}) and exhibited on Fig.\
9 for the dispherical particle the loci of transit times for
$t=0,0.002,0.02,0.04,0.06$. The loci for $t=0$ are points on Fig.\ 9 on the
principal ellipsoidal axis ($\xi=1$) at $\eta = 0,2 \pi /ka,4 \pi /ka, 1$ or $\eta
\approx 0,0.413,0.827,1$. The two coherent secondary sources at $(1,\pm1)$
nonlocally induce coherent tertiary focal points approximately at $(1,0)$,\ $(1,\pm
0.413)$ and $(1,\pm 0.827)$ which are focal points of the ``confined" trajectories
as exhibited by Fig.\ 6.  The crossing or tangency of trajectories at these induced
foci at the same time by Figs.\ 6 and 9 in the trajectory representation is, in
contrast, forbidden in Bohmian mechanics.$^{(\ref{bib:ph},\ref{bib:zm})}$.

Comparing Figs.\ 5 through 8 with Fig.\ 9, one sees that certain trajectories
transverse across some loci of transit time many times --- in alternating forward or
retrograde motion for a particular trajectory.  These multiple crossings imply that
a particle may be simultaneously at multiple locations manifesting strong
nonlocality. Thus the self-entangled dispherical particle ia not necessarily a point
particle, and its trajectory is that for a distributed particle.

For the ``confined" trajectories, while the series of trajectory segments that
alternate forward and retrograde motion render nil transit times from the
originating secondary source to the induced tertiary focal points, $t=0$, the
transit time from the originating secondary source to a non-focal point on the
trajectory in the same hemisphere of the originating secondary source is never
negative, i.e., $t \not< 0$; but for opposite hemispheres, never positive, $t \not>
0$.  On the other hand for ``free" trajectories, $t > 0$.

We note that the loci for transit times exhibit piecewise separation on Fig.\ 9. The
cut between $\eta = 1$ and $\eta = 4\pi/ka \approx 0.827$ in the locus for $t=0.002$
has closed for the locus for time $t=0.02$.  The cut centered at $\eta = 3\pi/ka
\approx 0.620$ for loci for times $t=0.002,0.02$ have closed for the locus for time
$t=0.04$.  The cut at $\eta = \pi/ka \approx 0.207$ would close for about transit
times $t > 0.156$, which are not exhibited on Fig.\ 9.

The equation of motion, Eq.\ (\ref{eq:eom}) is antisymmetric with regard to $\eta$.
Physically, a ``confined" trajectory, which has an inflexion point in Euclidean
3-space at the origin with the azimuthal coordinate $\phi$ changed by $\pm \pi$ so
that it transits the contours (disjointed and otherwise) of constant reduced action
in the upper and lower hemispheres in opposite directions. What is either forward or
retrograde motion in one hemisphere becomes reversed in the other hemisphere for
``confined" trajectories.  To generalize, the time of transit for any ``confined"
trajectories between $(\xi_0,\eta_0)$ and $(\xi_0,-\eta_0)$ is nil.  It follows that
the quantum dispherical particle has the same transit time from either secondary
point source to any other point on the ``confined" trajectory.

Contrasting Figs.\ 1, 2 and 4 with Fig.\ 9, one sees that shapes and connectivities
of the contours of reduced action and loci of transit times do not mimic each other
in the near region.  Note that the induced tertiary sources are neither intuitively
manifested by the contours for reduced action in Figs.\ 1, 2 and 4 nor obvious in
Eq.\ (\ref{eq:wd}).

After a ``confined" trajectory passes through the origin, it then, by the symmetry
of Fig.\ 1, transits the contours of constant reduced action in a reverse direction
from the direction that it had transited the symmetrically corresponding contours
before passing through the origin (that is a ``confined" trajectory transits
contours of constant $W_d$ and the loci of transit time in one hemisphere in the
reverse order for which it transited them in the other hemisphere). Since the origin
in Euclidean 3-space is an inflexion point and not a turning point for a ``confined"
trajectory as previously noted, those trajectory segments of the transit between the
upper secondary point source $(\xi,\eta)=(1,1)$ and the origin $(\xi,\eta)=(1,0)$
that are in the forward direction become retrograde with regard to time in the
transit from the origin $(\xi,\eta)=(1,0)$ and the lower secondary point source
$(\xi,\eta)=(1,-1)$ while those segments that are retrograde between the upper
secondary point source and origin become forward between the origin and lower
secondary point source.

\section{\itshape WELCHER WEG}

The secondary point sources of the trajectories for the quantum Young's diffraction
are the ellipsoidal focus points for prolate spheroidal coordinate system, which are
regular singular point in $\xi$ and $\eta$.$^{(\ref{bib:mf2})}$ The trajectory
equation, Eq.\ (\ref{eq:te3}) has branch point singularities in $\xi$ and $\eta$ at
these particular focus (secondary source) points. A trajectory for a self-entangled
dispherical particle beginning from from the lower secondary point source,
$(\xi,\eta)=(1,-1)$, with constant of the motion $0 <\eta_a \le 1$ propagates as a
``confined" trajectory with monotonically increasing $\eta$ from an initial value of
$-1$ until it reaches the value of $+1$ at the upper secondary point source
$(\xi,\eta)=(1,1)$, with nil transit time as previously shown.  At the upper
secondary point source, the ``confined" trajectory transitions while rounding the
branch point singularity at $(\xi,\eta)=(1,1)$ to become a ``free" trajectory with
the very same constant of the motion $\eta_a$. This transition is accompanied by a
change in the azimuthal coordinate $\phi$ of $\pm \pi$.  Along the ``free" segment
of the trajectory, $\eta$ now monotonically decreases from $+1$ to asymptotically
approaching the value of the trajectory's constant of the motion, $\eta_a$.
Analogously, any ``free" trajectory of the upper hemisphere may be coupled at
$(\xi,\eta)=(1,1)$ to the corresponding ``confined" trajectory with the same
constant of the motion, $\eta_a$.  An analogous situation exists for constants of
the motion $-1 \le \eta_a <0$ but with the roles of the upper and lower secondary
point sources reversed.

{\it Welcher Weg}?  Both ways concurrently.  The trajectory for the quantum
dispherical particle through any point in space will have at least a ``confined"
segment whose terminals are the two coherent secondary point sources.  The time of
transit for the quantum dispherical particle from either secondary point source to
any point on its trajectory will be the same.  As previously shown, the ``confined"
and ``free" trajectories span Euclidean 3- space.

Perhaps, quantum mechanics for over eight decades had been asking the wrong
question, ``{\it Welcher Weg}?" It should have been asking, ``How both ways
simultaneously?"

\appendix

\section*{APPENDIX. QUANTUM ERASURE}

Let us consider a hypothetical experiment for a quantum particle that is half
Young's diffraction and half Lloyd's mirror.  We combine these two experiments out
of phase in such a manner to swap quantum information on self-entangled dispherical
wave function for quantum information on spherical wave functions. Such a swapping
implies a quantum erasure. Herein, we are not studying the phenomenon of the quantum
erasure --- merely applying it as a confidence building measure to demonstrate that
a spherical wave function can be synthesized from two dispherical wave functions and
substantiate that nonlocal entangled quantum particles represented by dispherical
wave functions are physical.

Let us modify our hypothetical experiment of interference between two coherent
secondary point sources.  We still consider the behavior of a solitary quantum
particle. The first modification to Young's diffraction experiment is the
interference between two secondary point sources is still coherent but now
anti-correlated, that is the resultant quantum dispherical wave function $\psi_Y$
for Young's diffraction is now given as $\psi_Y = \psi_1-\psi_2$.  A sole primary
source emits a solitary quantum particle with a specified de Broglie wavelength that
after going through an initial 50:50 splitter actuates the secondary point sources 1
and 2. Secondary point source 1 is arbitrarily made the lower point source;
secondary point source 2, the upper secondary point source. Anti-correlation is
achieved by making the path length from the primary source to lower secondary point
source $N$ de Broglie wavelengths while inserting a second 50:50 splitter that in
turn splits the path from the primary source to the upper secondary point source
into two branches. One branch is $N-1/2$ de Broglie wavelengths long; the other,
$N+1/2$ de Broglie wave lengths. Having two branches to the upper secondary source
confounds a time-of-arrival analysis for determining {\it welcher Weg} for a
solitary quantum particle.

The next modification is to insert a half-silvered mirror along the the $\eta=0$
plane.  This produces two partial Lloyd's mirror experiments: one for the lower
secondary point source for $-1\le \eta \le 0$; the other for the upper secondary
point source for $0<\eta<+1$.  Concurrently the half-silvered mirror posits the
virtual point source for each Lloyd's mirror at the alternate secondary point
source.  Hence, the dispherical wave function for Lloyd's mirror in the lower
infinite hemisphere is given by $\psi_L=+(\psi_1+\psi_2)/2^{1/2},\ -1\le \eta\le 0$
and in the upper hemisphere is given by $\psi_L=-(\psi_1+\psi_2)/2^{1/2},\ -1\le
\eta\le 0$.

The half-silvered mirror also reduces the amplitude of dispherical wave function for
Young's diffraction by the factor $2^{-1/2}$ so that
$\psi_Y=+(\psi_1-\psi_2)/2^{1/2}$  throughout all space.  There exists interference
between the two dispherical waves, $\psi_Y$ and $\psi_L$.  These two dispherical
waves can be summed as

\[
\psi_Y + \psi_L =\left\{\begin{array}{cc}
                        2^{1/2}\psi_1, & \ \ -1\le \eta\le 0 \\
                        -2^{1/2}\psi_2, & \ \ 0\le \eta\le +1.
                        \end{array}
                \right.
\]

Hence, a spherical wave can be synthesized from two dispherical waves.  Cascaded
entanglement recovers the spherical wave. Both the spherical and dispherical waves
are wave functions of the Schr\"{o}dinger equation. It is just as valid to work with
dispherical wave functions as it is to work with spherical waves by the
superpositional principle of linear homogeneous differential equations.

\bigskip

\noindent {\bf Acknowledgements}

\bigskip

I heartily thank Marco Matone for his interesting discussions.  I also thank Robert
Carroll, Alon E. Faraggi, Bill Poirier and Robert E. Wyatt for their contributions
and encouragement.

\noindent {\bf References}

\begin{enumerate}\itemsep -.06in

\item \label{bib:prd34} E.\ R.\ Floyd, {\it Phys.\ Rev.}\ {\bf D 34}, 3246 (1986).

\item \label{bib:vigsym3} E.\ R.\ Floyd, {\it Gravitation and Cosmology:  From the
Hubble Radius to the Planck Scale; Proceedings of a Symposium in Honour of the
80$^{th}$ Birthday of Jean-Pierre Vigier}, ed.\ by R.\ L.\ Amoroso, G.\ Hunter, M.\
Kafatos and J.-P.\ Vigier, (Kluwer Academic, Dordrecht, 2002), extended version
promulgated as quant-ph/00009070.

\item \label{bib:irt} E.\ R.\ Floyd, {\it Found.\ Phys.}\ {\bf 37}, 1386 (2007),
quant-ph/0605120.

\item \label{bib:fm2} A.\ E.\ Faraggi and M.\ Matone, {\it Int.\ J.\ Mod.\ Phys.}\
{\bf A 15}, 1869 (2000), hep-th/98090127.

\item \label{bib:bfm} G.\ Bertoldi, A.\ E.\ Faraggi and M.\ Matone, {\it Class.\
Quant.\ Grav.}\ {\bf 17} 3965 (2000), hep-th/9909201.

\item \label{bib:rc} R.\ Carroll, {\it Can.\ J.\ Phys.} {\bf 77}, 319 (1999),
quant-ph/9904081; {\it Quantum Theory, Deformation and Integrability} (Esevier,
2000, Amsterdam) pp. 50--56; {\it Uncertainty, Trajectories, and Duality},
quant-ph/0309023.

\item \label{bib:mf} P.\ M.\ Morse and H. Feshbach, {\it Methods of Theoretical
Physics}, Part II (McGraw-hill, New York, 1953), p. 1284.

\item \label{bib:pdh} C.\ Philippidis, C.\ Dewdney and B.\ J.\ Hiley, {\it Neuvo
Cimento} {\bf 52B}, 15 (1979).

\item \label{bib:guantes} R.\ Guantes, A.\ S.\ Sanz, J.\ Margalef-Roig and S.\
Miret-Art\'{e}s. {\it Surf. Sci. Rep.} {\bf 53} 199 (2004).

\item \label{bib:bohm} D.\ Bohm, {\it Phys.\ Rev.}\ {\bf 85} 166 (1953).

\item \label{bib:prd26} E.\ R.\ Floyd, {\it Phys.\ Rev.}\ {\bf D 26}, 1339 (1982).

\item \label{bib:goldstein} H.\ Goldstein, {\it Classical Mechanics} 2$^{nd}$ ed.
(Addison-Wesley, Reading, 1980) p. 441.

\item \label{bib:mf2} P.\ M.\ Morse and H. Feshbach, {\it loc.\ cit.\ } Part I, p.
661.

 \item \label{bib:ph} P.\ R.\ Holland, {\it the Quantum Theory of Motion}
(Cambridge, Cambridge, 1993) pp. 85--86, 183, 201.

\item \label{bib:zm} Y.\ Zhao and N.\ Makri, {\it J.\ Chem.\ Phys.}\ {\bf 119}, 60
(2003)

\end{enumerate}

\noindent {\bf Figure Captions}

\bigskip

\noindent Fig.\ 1.  Contours of constant reduced action for $W_d =
0.5h,1h,1.5h,\cdots,5h$ for the quantum dispherical particle with  $m=1,\ \hbar=1,
a=1$ and $k=15.2$.

\noindent Fig.\ 2.  Contours of constant reduced action for the quantum dispherical
particle as solid lines. The dashed lines are the projection onto the
$\xi,\eta$-plane of the hyperboloids $\eta = 0.207, 0.620$ near where maximum
destructive interference occurs between $\psi_1$ and $\psi_2$.

\noindent Fig.\ 3.  Contours of constant reduced action for $W_d = 4h,5h,6h$ for the
quantum dispherical particle with $m=1,\ \hbar=1, a=1$ and $k=24.3$.  The dashed
lines are the projection onto the $\xi,\eta$-plane of the hyperboloids $\eta =
0.129, 0.388, 0.646, 0.905$ near where maximum destructive interference occurs
between $\psi_1$ and $\psi_2$.

\noindent Fig.\ 4.  Contours of constant reduced action for $W_d =
1.315815h,1.316815h,1.317815h$ in the vicinity of the origin for the quantum
dispherical particle with $m=1,\ \hbar=1, a=1$ and $k=15.2$.

\noindent Fig.\ 5.  Trajectory for the quantum dispherical particle originating from
the upper secondary source with constant of the motion $\eta_a = -\sin(\pi/18)$.
Note change of scale in $\rho$ by a factor of ten at $\rho=0.1$, denoted by the
dashed vertical line, to facilitate exposition.

\noindent Fig.\ 6.  A set of selected ``confined" trajectories for the quantum
dispherical particle from the upper secondary source for the constants of the motion
$\eta_a \approx -0,-0.383,-0.707,-0.924$.  A fifth trajectory for $\eta_a = -1$
superimposes upon the line $\xi=1$ in the range $0\le \eta \le 1$.

\noindent Fig.\ 7.  The trajectory for the quantum dispherical particle originating
from the upper secondary source with constant of the motion $\eta_a =
+\sin(\pi/32)$.

\noindent Fig.\ 8.  A set of selected ``free" trajectories for the quantum
dispherical particle from the upper secondary source for the constants of the motion
$\eta_a \approx +0,+0.383,+0.707,+0.924$.  A fifth trajectory for $\eta_a = +1$
superimposes upon the line $\eta=1$ in the range $1\le \xi $.

\noindent Fig.\ 9.  Loci of selected transit times.  At transit time $t=0$., induced
tertiary sources are at $(\xi,\eta) \approx (1,0.827),(1,0.413),(1,0)$.

\end{document}